\documentclass{iopart}

\usepackage{epsfig}

\begin{document}

\title{Hadron Correlations from Recombination and Fragmentation}

\author{Rainer J Fries}
\address{School of Physics and Astronomy, University of Minnesota, \\
Minneapolis, MN 55455}
\ead{fries@physics.umn.edu}

\begin{abstract}
We review the formalism of quark recombination applied to the
hadronization of a quark gluon plasma. Evidence in favor of the quark
recombination model is outlined. Recent work on parton correlations, 
leading to detectable correlations between hadrons, is discussed. 
Hot spots from completely quenched jets are a likely source of such 
correlations which appear to be jet-like.
It will be discussed how such a picture compares with measurement
of associated hadron yields at RHIC.
\end{abstract}

\submitto{JPG}

In the first part of this manuscript we revisit some basic arguments leading
to the formulation of recombination. Applications to RHIC are highlighted,
in particular the discovery of the quark counting rule for elliptic flow.
Toward the end of the first part we discuss some topics fundamental to 
recombination that have been disputed recently. We refer to \cite{Fries:04qm}
and the original papers cited therein for a more thorough review.

In the second part we introduce a recent extension of the recombination 
formalism that permits correlations between partons \cite{FMB:04}. 
These naturally lead to correlations between hadrons. We argue how jet-like
correlations at intermediate $P_T$ can be understood as signals of
quenched jets.

\section{Recombination of Quarks: a Review}

\subsection{Introduction}

In high energy collisions of particles or nuclei, partons are scattered out of
the bound states and new partons are created. The confinement property
of quantum chromodynamics (QCD) does not permit the existence of these colored 
partons as freely propagating states. Instead they have to be grouped into 
color singlet bound states that eventually become hadrons. 
This hadronization is a non-perturbative process in QCD and its
dynamics is not fully understood.

It has been proven that for a perturbative process at asymptotically large 
momentum transfer, the non-perturbative hadronization process can be 
separated from the hard scattering amplitude in a well-defined way. 
Hadronization in this limit is described by a set of fragmentation functions 
$D_{a/A}$ that parametrize, in a universal way, the probability that a hadron 
$A$ with momentum $P$ is created from a parton $a$ with momentum $p$ in the 
vacuum \cite{CoSo:81}.
Fragmentation functions have been measured in $e^++e^-$ collisions and
work well for hadron production at transverse momenta $P_T > 1$ GeV/$c$ in 
$p+p$ at RHIC energies.

However, the fragmentation process is not sufficient to explain hadron 
production in Au+Au collisions at RHIC at transverse momenta of several 
GeV/$c$. Two puzzling
observations lead to this conclusion. First, baryons are much more abundant
than predicted by fragmentation. PHENIX has observed a ratio of proton/pion
$\approx 1$ between 2 and 4 GeV/$c$ \cite{PHENIX:03ppi}. The value predicted 
by leading twist perturbative QCD and fragmentation is $\approx 0.2$ 
\cite{FMNB:03prc}. 
Second, protons and $\Lambda$s do not seem to suffer from jet quenching. 
The measured nuclear modification factors $R_{AA}$ are close to 1
\cite{PHENIX:03ppi,STAR:03llbar,STAR:03v2}, 
very unlike those of mesons which show 
strong jet quenching $R_{AA} <1 $ \cite{jetqu:03a}.

\subsection{Hadronization of Bulk Matter}

In central heavy ion collisions we create a hot and dense 
fireball consisting of deconfined quarks and gluons. It is apparent
that a hadronization picture that assumes a single parton in the
vacuum has to fail in such an environment. Hadronization involving two or
more partons can be described by higher twist fragmentation functions.
However, nothing is known about these non-perturbative quantities. 

Instead we can look at the extreme case of a phase space densely populated
with partons. In the vacuum, a single parton has to radiate gluons which
subsequently split into quark-antiquark pairs. Hadronization can only start 
once sufficient quark-antiquark pairs are available to turn into valence
quarks of hadrons. In a dense medium, enough partons are already there and
could just recombine or coalesce into hadrons: three quarks into a baryon, 
a quark-antiquark pair into a meson.

Recombination as a model of hadronization has a long history, see 
\cite{Fries:04qm} for some references.
Recombination of beam fragments \cite{DasHwa:77} has been experimentally 
confirmed by the observation of the leading particle effect in fixed target 
experiments \cite{WA82:93lp}.
The application to heavy ion collisions has been revived recently 
\cite{FMNB:03prl,FMNB:03prc,GreKoLe:03prl,HwaYa:02}
and is a center of great interest since then.

The yield of hadrons from a given parton system can be calculated starting
from a convolution of Wigner functions \cite{FMNB:03prc}. 
For a meson with valence (anti)quarks $a$ and $b$ we have
\begin{equation}
  \label{eq:reco}
  \fl
  \frac{d^3 N_M}{d^3 P}= \sum_{a,b} \int\frac{d^3 R}{(2\pi)^3} 
  \int\frac{d^3 qd^3 r}{(2\pi)^3} W_{ab}\left(\mathbf{R}-
  \frac{\mathbf{r}}{2},\frac{\mathbf{P}}{2}-\mathbf{q}; \mathbf{R}+
  \frac{\mathbf{r}}{2},\frac{\mathbf{P}}{2}+\mathbf{q} \right)
  \Phi_M (\mathbf{r},\mathbf{q}).
\end{equation}
Here $W_{ab}$ is the 2-particle Wigner function for partons $a$, $b$ and 
$\Phi_M$ is the Wigner function of the meson. The sum runs over all
possible parton quantum numbers. We note that recombination is dominated
by the lowest state in the Fock expansion of the hadron, i.e.\ the valence
quarks. Sea quarks and gluons do not seem to play a role, as can be inferred
from the quark counting rule for elliptic flow (see below). For practical
purposes the parton Wigner function is usually approximated by a product
of single particle phase space distributions $W_{ab}=w_a w_b$.
Several implementations of recombination have been discussed in the literature
\cite{FMNB:03prc,GreKoLe:03prl,HwaYa:02}, see \cite{Fries:04qm} for a review.

In order to obtain an estimate when recombination is important as a
hadronization mechanism, one can compare the yield of fragmentation 
and recombination starting from different parton spectra. The result of this 
competition is rather surprising. Thermal, or at least exponential, 
parton spectra
$w\sim e^{-P/T}$ play a special role. Recombination on such a parton
distribution leads to an exponential hadron distribution with the same
slope since
\begin{equation}
  w_a w_b \sim e^{-xP/T} e^{-(1-x)P/T} = e^{-P/T}.
\end{equation}
where $x$ gives the momentum fraction of parton $a$. 
Therefore recombination is more effective than fragmentation on any 
thermalized parton ensemble. On the other hand one can show that a power-law
parton spectrum favors fragmentation at least for large $P_T$, in accordance
with perturbative QCD.

Eq.\ (\ref{eq:reco}) does preserve momentum, but not energy. Therefore it
can only be safely applied in a kinematic region where mass effects are
small, i.e.\ for $P_T \gg M$. On the other hand we know that pQCD
fragmentation starts to dominate at very high $P_T$. The large jet quenching
at RHIC suppresses the contribution from fragmentation, so that recombination
effects can be observed at intermediate $P_T$.

\subsection{Recombination and RHIC data}

\begin{figure}
\begin{center}
\epsfig{file=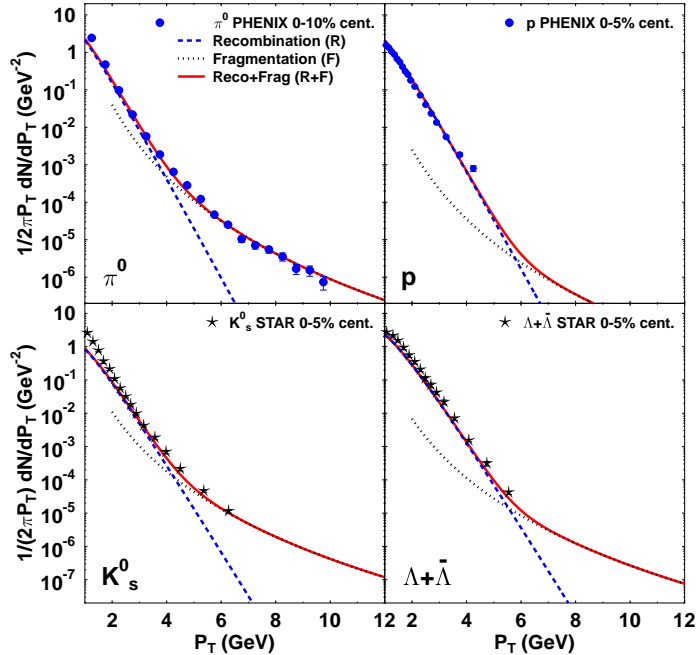,width=10cm}
\caption{\label{fig:spectra} Spectra of $\pi^0$, $p$, $K_0^s$ and $\Lambda+\bar
  \Lambda$ as a function of $P_T$ at midrapidity in central Au+Au collisions
  at $\sqrt{s}=200$ GeV \cite{FMNB:03prc}. Dashed lines are hadrons from
  recombination of the thermal phase, dotted line is pQCD with energy loss,
  solid line is the sum of both contributions. Data are from PHENIX ($\pi^0$, 
  $p$) \cite{PHENIX:03pi0,PHENIX:03ppi} and STAR ($K_0^s$, 
  $\Lambda+\bar\Lambda$) \cite{STAR:03llbar}.}
\end{center}
\end{figure}

Calculations for Au+Au collisions at RHIC energies assume a thermalized 
system of constituent quarks with a temperature $T$ around the phase 
transition temperature of the quark gluon plasma and strong radial flow.
Recombination is applied to this phase. To describe the high-$P_T$
spectrum of hadrons this has to be supplemented by a pQCD calculation using
fragmentation and taking into account energy loss \cite{FMNB:03prl,FMNB:03prc}.
Alternatively one can use another model to describe the power-law tail of the
$P_T$ spectrum, e.g.\ recombination of hard partons or soft-hard recombination
\cite{GreKoLe:03prl,HwaYa:03}.

Fig.\ \ref{fig:spectra} shows the $P_T$ spectrum of $\pi^0$, $p$, $K_0^s$ and 
$\Lambda+\bar\Lambda$ in central Au+Au collisions obtained in 
\cite{FMNB:03prc} with a model including recombination and fragmentation. 
The agreement with available data is excellent for $P_T > 2$ GeV/$c$. We note
that the hadron spectra exhibit an exponential shape up to about 4 GeV/$c$ 
for mesons and up to about 6 GeV/$c$ for baryons, where recombination of
thermal quarks dominates. Above, the spectra follow a power-law and production
is dominated by fragmentation.

\begin{figure}
\begin{center}
\epsfig{file=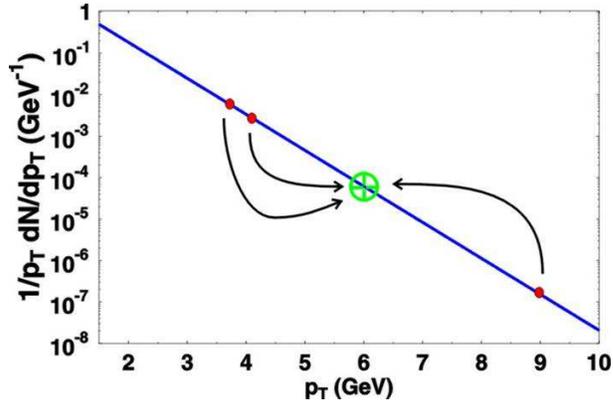,width=8cm}
\caption{\label{fig:sketch} Schematic action of recombination and 
  fragmentation on a transverse momentum spectrum of partons (solid line). 
  To produce a meson with $P_T=6$ GeV/$c$, the fragmentation process starts 
  with a parton having much more transverse momentum. On the other hand, 
  recombination can work with two coalescing partons having roughly 
  3 GeV/$c$ each.}
\end{center}
\end{figure}

Recombination naturally leads to a ratio $p/\pi \approx 1$. In the limit
$P_T \to \infty$ where masses are negligible and kinematics
is collinear, recombination of a thermal ensemble of partons exactly equals 
the statistical model of hadrons produced in thermal equilibrium. Therefore,
recombination solves the RHIC puzzle of hadron enhancement by introducing 
a mechanism that can produce hadrons close to thermal equilibrium. It also
pushes up the $P_T$ of hadrons, see Fig. \ref{fig:sketch}. The domination 
of recombination up to 6 GeV/$c$ is also the reason why jet quenching seems
to be absent in nuclear modification factors measured for baryons.
Recombination predicts a sharp decrease in $p/\pi$ beyond 4 GeV/$c$ which 
has not yet been seen in the data.

\subsection{Elliptic Flow}

Let us assume the parton phase exhibits elliptic flow $v_2^{\mathrm p}(p_T)$. 
Recombination makes a prediction for elliptic flow of any hadron species
after recombination \cite{Voloshin:02,FMNB:03prc}:
\begin{equation}
  \label{eq:v2}
  v_2(P_T) = n v_2^{\mathrm p}(P_T/n).
\end{equation}
Here $n$ is the number of valence quarks for the hadron. Hence recombination
predicts another striking difference between mesons (scaling with $n=2$) and 
baryons (scaling with $n=3$). Different mesons should follow the
same scaling law, even if the masses of the mesons are very different.

\begin{figure}
\begin{center}
\epsfig{file=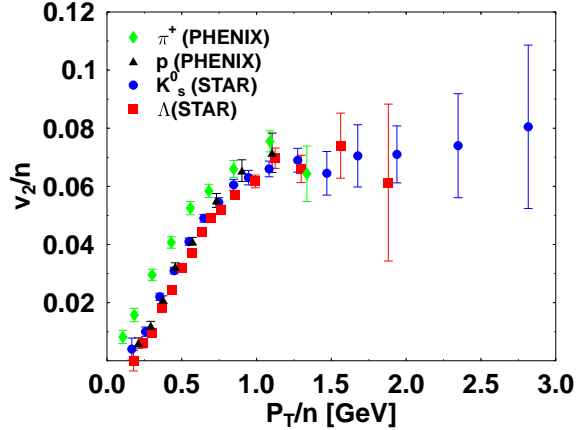,width=8cm}
\caption{\label{fig:v2} Elliptic flow $v_2$ for $\pi^+$, $p$, $K_0^s$ and
$\Lambda$ as a function of $P_T$ scaled by the number of valence quarks $n$ 
vs $P_T/n$. The data follows a universal curve, impressively confirming the 
quark scaling law predicted be recombination. Deviations for the pions are 
discussed in the text. Data are taken from PHENIX 
($\pi^+$, $p$) \cite{PHENIX:03v2} and STAR ($K_0^s$, $\Lambda$) 
\cite{STAR:03v2}.}
\end{center}
\end{figure}

Fig.\ \ref{fig:v2} shows the measured elliptic flow $v_2$ for several
hadron species in a plot with scaled axes $v_2/n$ vs $P_T/n$. All data
points (with exception of the pions) fall on one universal curve.
This is a an impressive confirmation of the quark scaling rule and the 
entire recombination picture. The pions are shifted to lower $P_T$, 
because most pions in the detectors, even at intermediate $P_T$,
are not from hadronization, but from secondary decays of hadrons. It was
shown recently that inclusion of $\rho$ recombination with subsequent 
decay of the $\rho$ resonance into 2 pions can largely account for the
shift \cite{GreKo:04rho}. We note that the scaled elliptic flow in Fig.\ 
\ref{fig:v2}
is equivalent to the {\it parton} elliptic flow. This would be the first
direct observation of a non-trivial observable in the parton phase.
An an immediate consequence we conclude that strange quarks
have the same elliptic flow as light quarks.

The quark scaling law for elliptic flow is reminiscent of classical quark 
counting rules, like the famous 2/3 ratio of the $\pi+N$ and $N+N$ cross 
sections. 
Some decades ago, these counting rules were very important to convince us 
that quarks
are indeed real degrees of freedom in high energy scatterings. Now
we have discovered another quark counting rule at RHIC. We can rewrite
(\ref{eq:v2}) as
\begin{equation}
  \frac{v_2^M(2p)}{v_2^B(3p)} = \frac{2}{3}
\end{equation}
where the superscripts $M$ and $B$ stand for the elliptic flow of a meson and
a baryon respectively. This quark counting rule is a clear hint that 
quarks are relevant degrees of freedom at RHIC. 
Furthermore, elliptic flow is an observable to describe collective behavior.
This implies that we created bulk matter with subhadronic
degrees of freedom. Therefore, this discovery could be a very important step 
to prove the creation of a quark gluon plasma in Au+Au collisions at RHIC.

\subsection{Further Remarks}

Recombination has been proven to be a valid description of hadron production
in high energy heavy ion collisions. Hadron spectra, hadron ratios, nuclear
modification factors and elliptic flow at intermediate momenta can be 
explained. The predictive power of recombination
lies in the fact that a consistent description of all hadron species has
been achieved with only one universal parametrization of the parton phase
in terms of temperature, radial and elliptic flow and volume.
Let us emphasize that the parton phase is {\it input} in the 
calculation, not a prediction. However, when hadron observables measured at
RHIC are consistently described by one universal parametrization of the 
parton phase, confidence grows that the right parameters have been found.

As mentioned above, recombination does not work at low $P_T$. This is 
not a failure of recombination itself, but a flaw of our simple description, 
that does not ensure energy conservation. Another problem that one 
encounters at low $P_T$ is the requirement that entropy must not decrease 
during hadronization.
Moreover, we know that hadrons at low $P_T$ do not escape directly from 
hadronization, but they suffer multiple rescattering in the hadronic phase. 
Therefore, direct comparison with data at low $P_T$ is difficult.
Implementations of resonance production and decay into a recombination
calculation can improve the description at low $P_T$ 
\cite{GreKoLe:03prl,GreKo:04rho}. On the other hand, measurements of 
resonances can be used to determine the strength of interactions in the
hadronic phase \cite{NMABF:03}.

\subsection{Constituent Quarks?}

Let us take a closer look at the question what is actually recombining.
From the quark counting rule for elliptic flow we can infer that only the
valence structure of a hadron is relevant. Recombination
involves only the lowest Fock state with sea quarks and gluons not being
relevant. This seems to be at odds with our usual picture of a hadron.
But this is not true. 

In high energy physics our picture of a hadron is mostly defined by 
parton distributions (PDFs) as seen in deep inelastic scattering. PDFs tell us 
that there is an infinite number of sea quarks and gluons in a hadron 
at a given $Q^2$, where $Q$ is a perturbative scale.
Parton distributions refer to average parton configurations in a free hadron.

We know that the momentum carried by sea quarks and gluons decreases if $Q$
decreases. The evolution can not be continued quantitatively into the 
non-perturbative domain $Q< 1$ GeV/$c$. However, it is clear that for 
$Q \approx \Lambda_{\mathrm QCD}$ perturbative quarks and gluons are not
longer appropriate degrees of freedom. Instead, one has to change to a 
constituent quark picture. In this picture quarks are dressed and acquire a 
dynamical mass and higher Fock states are suppressed due to the large quark
mass.

The constituent quark picture is not in contradiction with perturbative QCD
because it lives in a domain where perturbation theory does not work.
By applying non-perturbative methods in QCD it has been shown that there
is a smooth transition between both pictures. 
Lattice QCD \cite{Bowman:04}, instanton models \cite{DiaPe:84} and solutions 
of the Dyson-Schwinger equation \cite{Roberts:00} provide consistent 
results of dynamical quark masses as a function of $Q$. 
These functions smoothly interpolate between current quark masses relevant 
for perturbative QCD at large $Q$ and constituent quark masses at low $Q$. 
This running of the mass is also related to chiral symmetry breaking 
\cite{Roberts:00}.

At the phase transition with a temperature around $ 175$ MeV, it can be
ruled out that the deconfined phase is a perturbative quark gluon plasma.
Instead, it is natural to have effective degrees of freedom with
dynamical masses. A description of recombination that is linked to the chiral
phase transition is not yet available, but should be a priority of future
research.

\section{Hadron Correlations}

Recombination is thought to be the dominant mechanism of hadronization
at intermediate $P_T$ between 2 and 5 GeV/$c$. It has been argued that 
jet-like correlations of hadrons --- observed in this momentum range --- 
are an experimental fact that contradicts this view. This important
challenge has only been addressed recently \cite{FMB:04}.

\subsection{Associated Yields}

The STAR and PHENIX collaborations at RHIC have published data on 
dihadron correlations \cite{STAR:02corr,PHENIX:04corr}. They detect a 
trigger hadron $A$ in a certain $P_T$
range and measure the yield of secondary hadrons $B$ in another $P_T$ range
as a function of the relative angle $\Delta \phi$. The 2-hadron yield
$d N_{AB}/d(\Delta\phi)$ includes an uncorrelated background 
$d(N_A N_B)/d(\Delta\phi)$. 
We define the background subtracted associated yield {\it per trigger
particle} as
\begin{equation}
  Y_{AB} (\Delta\phi) = \frac{1}{N_A} \left( \frac{dN_{AB}}{d(\Delta\phi)}
  - \frac{d(N_A N_B)}{d(\Delta\phi)} \right).
  \label{eq:assyield}
\end{equation}
It was found that the associated yield in Au+Au collisions at RHIC shows an
enhancement above background around $\Delta \phi=0$ which is approximately the 
same as that for $p+p$ collisions. A smaller but broader 
enhancement can be seen in $p+p$ around $\Delta\phi=\pi$, but it is absent in 
Au+Au collisions. 

In $p+p$ collisions these signals can be interpreted as correlated emissions 
from jet cones. The near side peak ($\Delta\phi\approx 0$) is coming from
hadrons formed in the same jet as the trigger hadron. The far side associated
yield is due to hadrons from the jet recoiling against the jet of the trigger
particle. It is washed out due to higher order corrections. The transverse 
momenta of both the trigger and associated particles are in the range of a few
GeV/$c$. Therefore, there is no doubt that the correlations observed in $p+p$
are due to jet fragmentation.

On the other hand, near side correlations are also observed in Au+Au
collisions, while far side correlations seem to be severely dampened 
\cite{STAR:02corr,PHENIX:04corr}. The absence of far side correlations
can be understood as a consequence of the strong jet quenching in Au+Au 
at RHIC energies. 
However, the size of the near side correlations seem to be at odds with 
our understanding that a large recombination contribution in the relevant 
$P_T$ range exists.

\subsection{Correlations from Recombination}

We have discussed in the first part how recombination was successfully 
applied to describe hadron spectra
and elliptic flow starting from assumptions about the parton phase at
hadronization. One crucial simplification that has been always implemented 
so far is a factorization of any $n$-parton Wigner function into a product
of independent single parton distributions
\begin{equation}
  W_{1,\ldots,n} = \prod_{i=1}^n w_i
  \label{eq:prod}
\end{equation}
By definition, this factorization does not permit any correlations between 
partons. Consequently, no hadron correlations can emerge via recombination.
It has to be emphasized that the above factorization was chosen for 
simplicity and it was justified because single inclusive hadron spectra 
could be described very well.

It has been proven in \cite{FMB:04} that modifications of (\ref{eq:prod}) 
including correlations between partons do lead to correlations between 
hadrons upon recombination without deteriorating the quality of the 
description of single hadron spectra. We will come back to this later.
First, let us ask the legitimate question why there should be any 
correlations in the parton phase. The absence of correlations would indicate 
complete thermalization. However, the degree of equilibration at the time
of recombination is not exactly known.

Let us discuss one likely source of jet-like correlations in the medium.
It has been found that the strong jet quenching observed at RHIC is due to
energy loss of high-$p_T$ partons in the medium. The energy loss is estimated
to be up to 14 GeV/fm for a 10 GeV parton \cite{Wang:04jq}. This means that
most jets apart from those close to the surface are completely stopped,
dumping their energy and momentum into a cell of about 1 fm$^3$ in the
restframe of the medium. This results in a dramatic local heating, creating 
an ultra hot spot in the fireball.
Moreover, the directional information of the jet is preserved. The subsequent
evolution of the fireball might lead to partial diffusion of the hot spot.
Nevertheless, residual correlations might still exist and mimic a jet, because
the momentum of the jet has been transfered to the medium.
If partons of such a hot spot are picked they
would contribute to a jet-like correlation. Also a parton from a partially 
quenched jet together with a parton from the hot spot created can be 
correlated.
The STAR Collaboration presented evidence that jet cones can not be seen
as independent from the medium. Instead it is important that jets and the
medium mutually influence each other \cite{STAR:04text}.

\subsection{Modeling Correlations}

For 2-meson production we have to consider the following possible scenarios:
(1) Both mesons come from fragmentation, either of the same or of two
different jets. In a straight forward notation we will call this process F-F. 
(2) Both mesons emerge
from recombination of soft partons, denoted by SS-SS (S stands for soft).
(3) While one meson recombines, the other emerges from the coalescence of
a soft parton and a hard parton (from a jet). The latter is also called 
soft-hard coalescence, so we call this process SH-SS.
(4) One meson comes from recombination while the other is from fragmentation
(F-SS). (5) Both mesons can come from soft-hard recombination (SH-SH).
(6) One meson fragments while the other is from soft-hard recombination
(F-SH). 

\begin{figure}
\begin{center}
\epsfig{file=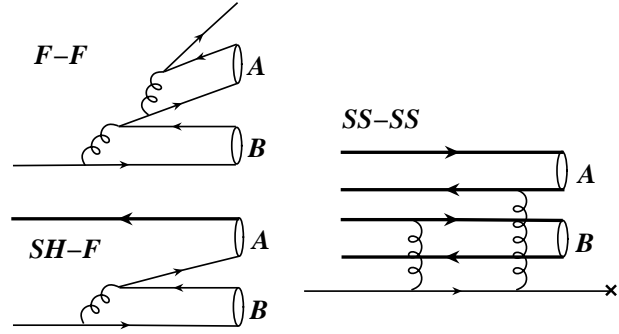,width=9cm}
\caption{\label{fig:2-meson} A schematic picture of correlated production of 
two mesons A and B through the processes F-F, F-SH and SS-SS. 
Thick lines denote soft partons.}
\end{center}
\end{figure}

All processes can produce both correlated and uncorrelated pairs. 
Here we are most interested in correlations coming from pure recombination
(SS-SS). However, conventional jet correlations (coming from F-F) have to be
taken into account when comparing to data. As an example for soft-hard
correlations we will also estimate correlations from F-SH.
These three processes are schematically depicted in Fig.\ \ref{fig:2-meson}.

\subsection{Dimeson Production from Soft Partons (SS-SS)}

We start with a simple extension of the correlation-free factorization
(\ref{eq:prod}). We want to assume that correlations are a small effect and
restrict ourselves to 2-particle correlations $C_{ij}$, so that a 4-parton
Wigner function can be written as
\begin{equation}
  W_{1234} \approx w_1 w_2 w_3 w_4 \big( 1 + \sum_{i<j}C_{ij} \big).
  \label{eq:newfact}	
\end{equation}
The correlation functions $C_{ij}$ between parton $i$ and parton $j$ are in 
principle of arbitrary shape, but we want to assume that they vary slowly 
with momentum and that they are only non-vanishing in a subvolume $V_c$ 
of the fireball.

A picture of hot spots created by quenched jets motivates the following 
Gaussian ansatz in rapidity and azimuthal angle
for the correlation functions:
\begin{eqnarray}
  \label{eq:partcorr}
  C_{ij}  &=&  c_0 \, S_0 \, f_0 
    e^{-(\phi_i-\phi_j)^2/(2 \phi_0^2)} \,
    e^{-(y_i-y_j)^2/(2y_0^2)} \\    \nonumber
   &+& c_\pi \, S_{\pi} \,f_{\pi}
     e^{-(\phi_i-\phi_j+\pi)^2/(2 \phi_\pi^2)} \,
    e^{-(y_i-y_j)^2/(2y_\pi^2)} . 
\end{eqnarray}
Here $\phi_{0,\pi}$ and $y_{0,\pi}$ are the widths of the Gaussians in 
azimuth and rapidity, respectively. The two terms of the sum correspond 
to correlations initiated by an energetic parton ($\phi=0$) and its recoil 
partner ($\phi = \pi$).  $c_0$ and $c_\pi$ give the strength of the near 
side and far side correlations, while the functions 
$f_{0,\pi}(p_{Ti},p_{Tj})$ describe the transverse momentum dependence of 
the correlations. The functions $S_{0,\pi}(\sigma_i,\sigma_j)$ parametrize 
the spatial localization of the parton correlations on the hypersurface 
$\Sigma$. For simplicity we assume that $S_{0,\pi} = 1$, if 
$\sigma_i,\sigma_j \in V_c$ and $S_{0,\pi}=0$ otherwise. 

The 2-meson yield is given by a convolution of the partonic Wigner function
$W_{1234}$ with the Wigner functions $\Phi_A$, $\Phi_B$ of the mesons with
an additional integration over the hadronization hypersurface $\Sigma$
\cite{FMB:04}
\begin{equation}
  E_A E_B \frac{d^6N_{AB}}{d^3 P_A d^3 P_B} = C_{AB} \int_\Sigma
  d \sigma \, \Phi_A \otimes W_{1234} \otimes \Phi_B.
  \label{eq:2-mes}
\end{equation}
We restrict our discussion to near side correlations. To this end we neglect
the second line in (\ref{eq:partcorr}). We also work in a fixed $P_T$ window
and will therefore neglect the transverse momentum dependence by setting
$f_0 =1$, absorbing the normalization into the constant $c_0$. With more data
available in small $P_T$ bins, the functional shape of $f_0$ and $f_\pi$
will be explored in the future. We assume $c_0 \ll 1$ which permits
omitting quadratic terms like $c_0^2$ or $c_0 v_2$.

Using Boltzmann distributions for the single parton distributions $w_i$ 
with temperature $T$, radial flow rapidity $\eta_T$, and a boost-invariant
hypersurface $\Sigma$ at proper time $\tau$, Eqs.\ (\ref{eq:2-mes}),
(\ref{eq:newfact}) and (\ref{eq:partcorr}) lead to
\begin{eqnarray} 
  \label{eq:tre}
  \fl
  \frac{d^6 N_{AB}}{P_{TA} dP_{TA} d\phi_A dy_A P_{TB} dP_{TB} d\phi_B dy_B} =
  \left( 1+2 \hat c_0 + 4 \hat c_0 e^{-(\Delta\phi)^2/(2\phi_0)^2} \right) \\
  \nonumber
  \fl \quad\times h_A(P_{TA}) \left(1+ 2v_{2A}(P_{TA})\cos(2\phi_A) \right)
  h_B(P_{TB}) \left(1+ 2v_{2B}(P_{TB})\cos(2\phi_B) \right). 
\end{eqnarray}
$v_{2i}$ is the elliptic flow coefficient for hadron $i$ and
$\Delta\phi=|\phi_A - \phi_B|$ is the relative azimuthal angle. The factor
$Q=4$ in front of the Gaussian term is called the amplification 
factor. We have introduced the short notation
\begin{equation}
  \fl h_i(P_T) = C_i \frac{\tau A_T}{(2\pi)^3} M_T I_0 \left(\frac{P_T \sinh 
  \eta_T}{T} \right) K_1 \left(\frac{\left(\sum_{j=1,2}\sqrt{m_j^2+P_T^2/4}
  \right)\cosh\eta_T}{T} \right)
\end{equation}
where $m_j$ are the masses of the recombining (anti)quarks, $M_T$ is the 
transverse mass of the meson, $A_T$ is the transverse area of the fireball
and $C_i$ is the degeneracy factor for meson $i$.
The integration over the correlation volume has been absorbed into the 
correlation strength leading to the new parameter $\hat c_0 \approx c_0 V_c/
(\tau A_T)$.

The single meson spectra take the simple form
\begin{equation}
  \frac{d^3 N_i}{P_{Ti} dP_{Ti}d\phi_i dy_i} = h_i(P_{Ti})(1+\hat c_0)
  \left( 1+2v_{2i}(P_{Ti})\cos(2\phi_i) \right).
  \label{eq:singlemes}
\end{equation}
We note that the only modification to the single meson spectrum --- compared 
to the case without correlations \cite{FMNB:03prc} --- 
consists of a moderate rescaling $1 \to 1+\hat c_0$. For small numerical 
values of $\hat c_0$ this can be easily absorbed in the overall 
normalization, so that a good description of single particle spectra 
in the recombination formalism can still be achieved with a consistent set
of parameters for the parton phase.

\subsection{Amplification of Correlations}

The factor $Q=4$ implies an enhancement of the correlations on the hadron
side compared to the parton side. The effect is essentially the same as in
the amplification of elliptic flow by the number $n$ of valence quarks in the
hadron. In the case of 2-parton correlations, $Q$ counts the number of 
possible correlated pairs between the $n_A$ (anti)quarks of meson $A$ and the
$n_B$ (anti)quarks of meson $B$. Since we work in a weak correlation limit 
where quadratic terms are suppressed, we do not take into account more than one
pair of correlated partons. Apparently
\begin{equation}
  Q=n_A n_B,
\end{equation}
thus $Q=6$ for a meson-baryon pair and $Q=9$ for a baryon-baryon pair.
The term $2 \hat c_0$ in Eq.\ (\ref{eq:tre}) comes from the possible 
correlation of the quark-antiquark pair inside of one of the two mesons.

We can now study the associated yield defined in Eq.\ (\ref{eq:assyield}), 
in a certain kinematic window,
by subtracting the uncorrelated yield $N_A N_B$, determined by 
(\ref{eq:singlemes}),
from the total dimeson yield given in (\ref{eq:tre}). The terms proportional
to $1+2\hat c_0$ cancel and hence
\begin{equation}
  2\pi N_A Y_{AB} (\Delta\Phi) = Q \hat c_0 e^{-(\Delta\phi)^2/(2\phi_0)^2}
  N_A N_B.
  \label{eq:res}
\end{equation}
The $N_i$ are single particle yields in the kinematic window of the trigger 
meson or associated meson
\begin{equation}
  N_i = 2 \pi \int dy_i \int dP_T P_T h_i(P_{Ti}).
\end{equation}
Correlations in rapidity have not been studied here. Since we integrate over
the rapidities any residual effects can be absorbed into the constant 
$\hat c_0$. We note that the result in (\ref{eq:res}) also holds for
meson-baryon and baryon-baryon correlations with the appropriate 
amplification factor $Q$.

It is also interesting to know the size of the subtracted background. It
is given by
\begin{equation}
  \frac{2\pi}{N_A} \frac{d(N_A N_B)}{d (\Delta\Phi)} =
  N_A N_B \left(1+2\hat c_0 + 2 \bar v_{2A} \bar v_{2B} \cos(2\Delta\phi)
  \right).
\end{equation}
$\bar v_{2i}$ is the elliptic flow of meson $i$ averaged over the respective
$P_T$ window.
We note that the background already has a non-trivial dependence on the
relative azimuthal angle coming from the the elliptic flow of the mesons.

\subsection{Correlations from Fragmentation}

Before comparing to data we have to find a way to describe the presence of
correlations from the F-F process. This requires a study in
the framework of dihadron fragmentation functions. These functions
are under investigation, but their functional form is still poorly known 
\cite{deFloVa:03}. Therefore, we choose a simple model based on single hadron
fragmentation functions. After a parton $a$ with momentum $p$ 
has fragmented into a hadron $A$ with momentum $P_A=z_A p$, described
by a fragmentation function $D_{a/A}(z_A)$, we assume that the production of
the second hadron $B$ with momentum $P_B$ is given by a fragmentation
function $D_{a/B}(z_B)$ where $P_B = z_B (1-z_A)p$. This model has clear
limitations, but it should give a rough estimate of the transverse momentum
dependencies.

Furthermore we introduce a Gaussian distribution of the relative azimuthal
angle and rapidity of the two hadrons $A$ and $B$. It can be shown that 
\begin{eqnarray}
  \fl \frac{d\left( N_A Y_{AB}(\Delta\phi)\right)}{dP_{TA} dP_{TB}} =& 2\pi I 
  (2\pi\phi_0^2)^{-1/2} e^{-(\Delta\phi)^2/(2\phi_0^2)} \sum_a \int_{z_0}^{z_1}
  \frac{d z_A}{z_A (1-z_A)} \nonumber\\ 
  &\times g_a \left( \frac{P_{TA}}{z_A} +\Delta E \right)
  D_{a/A}(z_A) D_{a/B}\left( \frac{z_A P_{TB}}{(1-z_A) P_{TA}} \right).
\end{eqnarray}
$g_a(p) = E_a d^3N_a/dp$ is the spectrum of the fragmenting parton $a$
and $I$ is the result of integrating the rapidities of the two hadrons 
in the kinematic window. $\phi_0$ is the width of the azimuthal correlation
and the limits for $z_A$ are given by $z_0=2P_{TA}/\sqrt{s}$ and
$z_1=P_{TA}/(P_{TA}+P_{TB})$. We also included a possible energy loss 
$\Delta E$ of parton $a$ before fragmentation.

The fragmentation from a single parton provides correlated hadrons. 
The uncorrelated background can be easily estimated from fragmentation of two 
hadrons from two independent jets.

\subsection{Correlations from Soft-Hard Processes}

Among the possible 2-hadron production mechanisms involving soft-hard
recombination, we choose the F-SH process as an example for our discussion. 
Soft-hard recombination is not part of the original recombination model
in \cite{FMNB:03prl,FMNB:03prc}, but it is taken into account by other
groups \cite{GreKoLe:03prl,HwaYa:03}.
The relative importance of soft-hard recombination is not clear, since
both approaches fit the experimental data well. It was argued that soft-hard
processes are a good source of hadron correlations. We want to shed some
light on this discussion.

We use the following model for dimeson production in the F-SH channel.
Suppose a hard parton $a$ with momentum $p_a$ produces an additional $b\bar b$
pair through branching, where $b$ is any light quark flavor. The pair 
$a\bar b$ hadronizes into a meson, while $b$ can pick up another (anti)quark
$c$ from the surrounding medium to coalesce into a meson $B$. 
The production of $A$ can be described by a fragmentation function 
$D_{a/A}(z_A)$ while we want to assume here that parton $b$ remains with 
momentum $(1-z_A)p_a$.

With a Boltzmann distribution for parton $c$ it follows that
\begin{eqnarray}
  \fl \frac{d\left( N_A Y_{AB}(\Delta\phi)\right)}{dP_{TA} dP_{TB}} =& 2\pi
  I \hat v \frac{8C_B M_{TB}}{P_{TB}} g_a(p_a) (2\pi\phi_0^2)^{-1/2} 
  e^{-(\Delta\phi)^2/(2\phi_0^2)}
  I_0 \left(\frac{P_{TB} \sinh \eta_T}{2T} \right) 
  \nonumber \\  & \times
  K_1 \left(\frac{\sqrt{m_c^2+P_{TB}^2/4}\cosh\eta_T}{T} \right)
  D_{a/A} \left( \frac{P_{TA}}{P_{TA}+P_{TB}/2} \right)
\end{eqnarray}
with $p_a = P_{TA} + P_{TB}/2 +\Delta E$. $\hat v <1$ is a normalization
constant arising from the restriction of the $\Sigma$ integration to a subspace
given by the intersection with a jet cone.

\subsection{Numerical Results}

For the numerical evaluation we use the set of parameters found in 
\cite{FMNB:03prc} to fit the single hadron spectra and elliptic flow
measured at RHIC. We the minijet calculation in \cite{FMS:02} and KKP 
fragmentation functions \cite{KKP:00}. We choose the windows 
2.5 GeV/$c$$\le P_{TA}\le$4.0 GeV/$c$ for trigger particles, 
1.7 GeV/$c$$\le P_{TB}\le$2.5 GeV/$c$ for associated particles and
$|y_A|$, $|y_B|<0.35$ as in the recent analysis of the PHENIX experiment 
\cite{PHENIX:04corr}. We fix the azimuthal correlation width to be
$\phi_0 = 0.2$ in all channels, in rough agreement with the experimental
data. In order to have a measure of the absolute strength of the correlation
we follow \cite{PHENIX:04corr} and integrate $Y_{AB}$ over the near-side 
peak
\begin{equation}
  Y_{AB}^{\mathrm{cone}} = \int_0^{0.94} d(\Delta\Phi) Y_{AB}(\Delta\Phi).
\end{equation}

\begin{figure}
\begin{center}
\epsfig{file=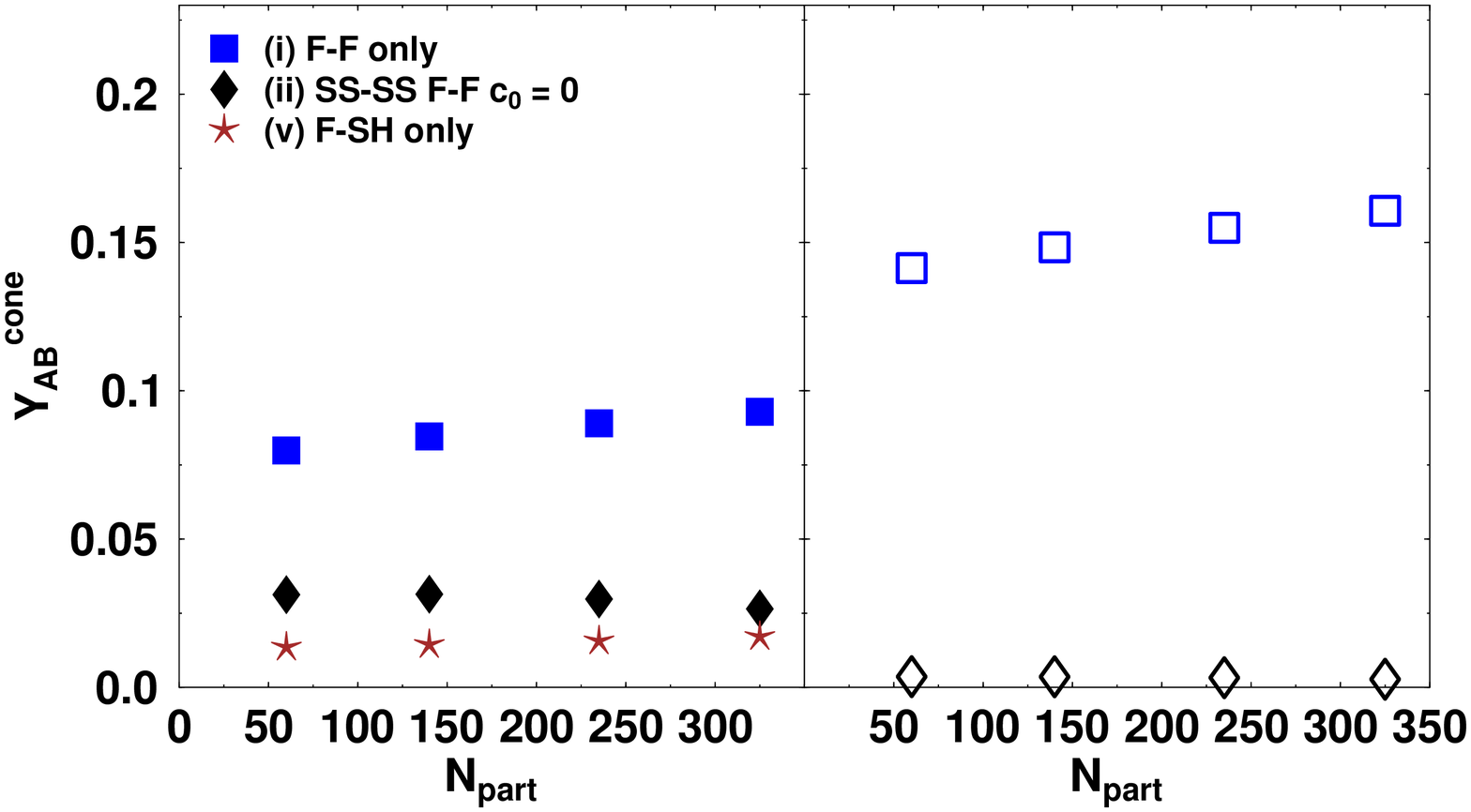,width=10cm} \\
\epsfig{file=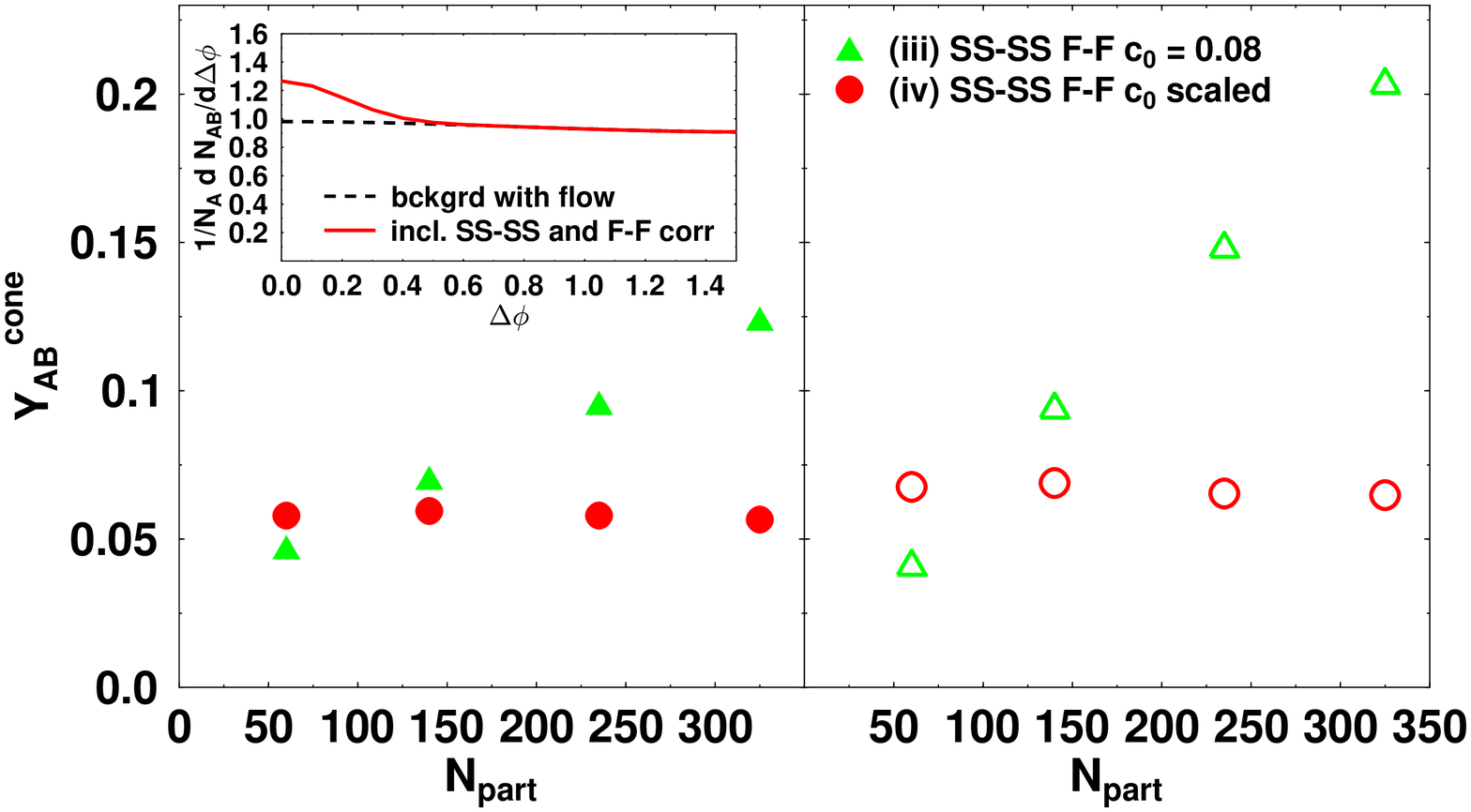,width=10cm}
\caption{\label{fig:res} The associated yield $Y_{AB}^{\mathrm{cone}}$ for
baryon triggers (right panels) and meson triggers (left panels) as a function
of $N_{\mathrm{part}}$. Squares are F-F only (i), diamonds are F-F and SS-SS
with $\hat c_0=0$ (ii), triangles are F-F and SS-SS with $\hat c_0=0.08$ (iii),
circles are F-FF and SS-SS with $\hat c_0=0.08\times 100/N_{\mathrm{part}}$ 
(iv), stars are F-SH with $\hat v=0.5$ [pions only](v). Insert: associated
yield $Y_{AB}$ as a function of $\Delta\phi$ before background subtraction
from F-F and SS-SS at impact parameter $b=8$ fm.}
\end{center}
\end{figure}

Fig.\ \ref{fig:res} shows the integrated associated yield of hadrons 
for the case that the trigger is a baryon (proton or antiproton) and 
a meson (pion or kaon) for different centralities. We consider the 
following cases: (i) F-F, i.e.\ fragmentation only. Our results are 
in rough agreement with the data without any adjustable parameter in our
model. This means that our model for double fragmentation would be
suitable to describe correlations in $p+p$. 
(ii) F-F and SS-SS with $\hat c_0 = 0$. Keeping fragmentation and turning on
recombination without correlations dilutes the signal as expected. 
The effect is particularly strong for baryon triggers. 
(iii) F-F and SS-SS with $\hat c_0 = 0.08$, and (iv) F-F and SS-SS with 
$\hat c_0 = 0.08 \times 100/N_{\mathrm{part}}$. We remember that $\hat c_0$
contains a space-time integral which can depend on centrality.
Therefore we test two different scenarios, $\hat c_0$ constant and
$\hat c_0$ scaling like $\sim N_{\mathrm{part}}^{-1}$. These cases correspond 
to the correlation volume increasing, $V_c \sim N_{\mathrm{part}}$, and $V_c$ 
being held constant respectively. The first case leads to a strong rise of 
$Y_{AB}^{\mathrm{cone}}$ with increasing centrality that is not observed in 
the data. The scenario of approximately
constant correlation volume $V_c$ leads to a nearly constant 
$Y_{AB}^{\mathrm{cone}}$ which is in qualitative agreement with PHENIX 
measurements \cite{PHENIX:04corr}.
(v) We also consider the case F-SH (for pions only). A choice of $\hat v =0.5$ 
for the parameter leads to sizable correlations. However, the yield of 
hadrons from the F-SH channel is small compared to that from F-F and SS-SS.
This is due to the small contribution from soft-hard recombination in the Duke
parametrization as already discussed in \cite{FMNB:03prc}.

We note that a realistic model of correlations from fragmentation and 
recombination can reproduce associated yields in qualitative agreement with
the data. The parameters of the parton phase can be chosen consistently so that
the excellent fits of single hadron spectra and elliptic flow are preserved.
Correlations on the parton level need only to be $~ 8\%$ in order to 
reproduce the data. Soft-hard recombination naturally leads to correlations
between hadrons, but this hadronization channel is not 
necessary to obtain a good description of the data. More details can be found
in \cite{FBM:04}

\section{Conclusions}

I have presented evidence that hadronization in heavy ion collisions is
dominated by recombination from a thermalized parton phase. A theoretical
description of recombination for $P_T>2$ GeV/$c$ is available and
naturally explains the enhanced baryon production at RHIC. The quark counting
rule for elliptic flow, impressively confirmed by experiment, might prove
to be an important cornerstone to make the case for the quark gluon plasma.

I also emphasized that our description of the dynamics of the recombination
process is still insufficient, which hampers progress in our understanding  
of low $P_T$ hadron production. In particular, we do not have a quantitative
theory that connects recombination to the chiral phase transition.

On the other hand, it has been shown that experimental findings of jet-like
hadron correlations at intermediate $P_T$ are not in contradiction with
the existence of recombination. Correlations among partons can be introduced 
into the existing formalism and naturally lead to correlations among hadrons.
Jet-like correlations in the medium can be created by jets that are quenched
by the medium and leave a hot spot.

\ack I would like to thank my coworkers B M\"uller, S Bass and C Nonaka. 
I also want to thank the organizers of Hot Quarks 2004 for a wonderful 
workshop. This work was supported by DOE grant DE-FG02-87ER40328.

\end{document}